\documentclass{elsart}

\usepackage{graphicx}
\usepackage{amssymb}

\begin{document}

\begin{frontmatter}



\title{Patterns of variability in gamma-ray blazars}


\author{Luigi Foschini}
\ead{foschini@iasfbo.inaf.it}

\address{INAF/IASF-Bologna, Via Gobetti 101, 40129 Bologna (Italy)}

\begin{abstract} 
Preliminary results of a systematic study on the simultaneous optical-to-X-rays variability in blazars are presented. Data from \emph{Swift} observations of four bright $\gamma-$ray blazars (3C~$279$, ON~$+231$, S5~$0716+71$, PKS~$2155-304$) have been analyzed, compared, and discussed. Specifically, 3C~$279$ shows a variable flattening in the low energy part of the X-ray spectrum that appears to be confined in a specific X-ray vs optical/UV fluxes region. Some implications are shortly analyzed.
\end{abstract}

\begin{keyword}
Blazars \sep Variability \sep Optical \sep Ultraviolet \sep X-rays
\end{keyword}

\end{frontmatter}

\section{Introduction}
Active galactic nuclei with relativistic jets pointed to or within a few degrees of the observer's direction are generally named blazars. However, this name collects a variety of phenomena, most of them are still to be understood. The spectral energy distribution (SED - $\nu$  vs $\nu F_{\nu}$) of these sources always displays a specific shape, which is the result of two types of emission: the first one, extending from radio to soft X-rays, is thought to be due to the synchrotron emission from relativistic electrons; while the second one, peaking at energies in the X-/$\gamma-$ray band, is generally attributed to inverse-Compton (IC) emission, resulting from the transfer of energy from the same population of relativistic electrons impinging on low-energy seed photons. The latter population of photons can be the same synchrotron radiation (Synchrotron Self-Compton, SSC) or a source external to the jet, like the accretion disk or the broad-line region (External Compton, EC). Fossati et al. (1998) and Ghisellini et al. (1998) suggested that blazars follow a trend –- \emph{the blazar sequence} -– where sources with low bolometric luminosity have both peaks at high frequencies (soft X-rays for synchrotron and TeV for IC), while, as the luminosity increases, both peaks shift to lower frequencies (optical/infrared for synchrotron and MeV/GeV for IC). See, however, Padovani (2007) and references therein for a discussion. Recent updates of the blazar sequence can be found in Ghisellini \& Tavecchio (2008) and Maraschi et al. (2008).

The shape and normalization of the SED during quiescence can be different when the source is in outburst and it is still to be understood how and when these changes occur. To investigate this problem, I started a study of the simultaneous optical/UV/X-rays variability of a sample of $\gamma-$ray blazars (detected both at MeV/GeV and TeV energies) to search for some common patterns of variability. Here I present the preliminary results.

\section{Sample selection and analysis}
Blazars are generally studied with multiwavelength (MW) campaigns, which are triggered by events of outburst and therefore the quiescence state is often neglected. However, to understand blazars, it is necessary to know how the source behaves during all its lifetime.  Therefore, I focus on bright blazars, which can be almost always observed by \emph{Swift} satellite (Gehrels et al. 2004) both during quiescence and outburst. It is also necessary to select blazars observed many times (to have a sufficient number of points) and with the same instruments settings (to have the same type of MW information). For the moment, I found sufficient data for the 4 blazars here shown, but work is in progress to enlarge the sample.

\emph{Swift} data have been processed, cleaned and analyzed by using the software package \texttt{HEASoft 6.5} and the calibration data base \texttt{CALDB} updated on June 25, 2008. Data from the X-Ray Telescope (XRT, $0.2-10$~keV, Burrows et al. 2005) were analyzed with the \texttt{xrtpipeline} task with standard parameters. XRT was set in photon counting (PC) mode for the observations of 3C~$279$, ON~$+231$, S5~$0716+714$ and single to quadruple pixel events were selected (grades $0-12$). In the case of PKS~$2155-304$, XRT was set in window timing (WT) mode for most of the pointings and events with grades $0-2$ were considered. A small numbers of observations in PC mode were performed: in this case, being the source count rate so high to be affected by pile-up, it was necessary to apply a correction by removing the central region of the point-spread function. Therefore, the source extraction region was an annulus with $10''$ and $50''$ of internal and external radius, respectively. The extracted spectra were rebinned to have at least $20-30$ counts per energy bin, depending on the available statistics, and fitted with models in \texttt{xspec v. 11.3.2ag}. I used a single redshifted power-law or a broken power-law models, with Galactic absorption from Kalberla et al. (2005). Sometimes, the low energy part of the X-ray spectrum (generally $\lesssim 2$~keV) shows a curvature due to a deficit of photons that cannot be fitted with a broken power-law model. In this case, the single power-law model is multiplied by an exponential roll-off in the form $\exp (-E_{f}^2/E^2)$, where $E_{f}$ is the folding energy of the roll-off. The fits are summarized in the Tables~\ref{tab:Summary1}, \ref{tab:Summary2}, \ref{tab:Summary3} and \ref{tab:Summary4}.

\begin{table}[!t]
\scriptsize
\caption{\scriptsize Summary of \emph{Swift}/XRT observations of 3C~$279$ ($z=0.5362$, $N_{\rm H, Gal}=2.05\times 10^{20}$~cm$^{-2}$, Kalberla et al. 2005). Columns: (1) Date of observation [DD-MM-YYYY]; (2) Photon index (for the power-law model) or soft photon index (for the broken power-law model); (3) Break energy $E_{b}$ for the broken power-law model or folding energy $E_{f}$ of the exponential roll-off for the single power-law model plus low energy photon deficit [keV]; (4) Hard photon index for the broken power-law model; (5) Model normalization [$10^{-3}$~ph~cm$^{-2}$~s$^{-1}$~keV$^{-1}$]; (6) Reduced $\chi^2$ and degrees of freedom of the spectral fit; (7) Notes.}
\centering
\bigskip
\begin{tabular}{rrrrrrr}
\hline
Date        & $\Gamma$       & $E_{b}/E_{f}$ & $\Gamma_2$ & $N$  & $\tilde{\chi}^2$/dof & Notes\\
(1)         & (2)            & (3)           & (4)        & (5)  & (6)                  & (7)\\
\hline
$13-01-2006$ & $1.53\pm 0.03$ & {}      & {}         & $3.8\pm 0.1$ & $1.10/123$ & {}\\
$14-01-2006$ & $1.58\pm 0.04$ & {}      & {}         & $3.7\pm 0.2$ & $0.97/74$  & {}\\
$16-01-2006$ & $1.5\pm 0.1$   & {}      & {}         & $3.2\pm 0.5$ & $1.26/9$   & {}\\
$17-01-2006$ & $1.28_{-0.08}^{+0.06}$ & $1.6_{-0.3}^{+0.2}$ & $1.74\pm 0.08$ & $2.14_{-0.06}^{+0.08}$ & $1.11/148$ & {}\\
$18-01-2006$ & $1.34_{-0.10}^{+0.07}$ & $1.6_{-0.4}^{+0.3}$ & $1.76_{-0.09}^{+0.10}$ & $2.41_{-0.08}^{+0.12}$ & $0.90/121$ & {}\\
$19-01-2006$ & $1.3_{-0.9}^{+0.1}$ & $1.6_{-1.0}^{+0.7}$ & $1.8\pm 0.2$ & $2.1_{-0.1}^{+1.6}$ & $1.04/34$ & {}\\
$12-01-2007$ & $1.60\pm 0.03$ & {}      & {}         & $6.4\pm 0.2$ & $1.47/111$ & {}\\
$17-01-2007$ & $1.47\pm 0.09$ & {}      & {}         & $5.8\pm 0.6$ & $0.86/19$  & {}\\
$18-01-2007$ & $1.6\pm 0.1$   & {}      & {}         & $5.3\pm 0.6$ & $0.94/16$  & {}\\
$19-01-2007$ & $1.7\pm 0.2$   & $0.36_{-0.16}^{+0.13}$ & {} & $6.6_{-1.4}^{+1.7}$ & $1.29/16$ & {}\\ 
$20-01-2007$ & $1.55\pm 0.07$ & {}      & {}         & $5.2\pm 0.4$ & $1.15/29$ & {}\\
$21-01-2007$ & $1.59\pm 0.07$ & {}      & {}         & $5.2\pm 0.4$ & $0.82/32$ & {}\\
$26-02-2007$ & $1.41\pm 0.09$ & {}      & {}         & $3.6\pm 0.4$ & $0.72/18$ & {}\\
$14-06-2007$ & $1.3_{-0.2}^{+0.1}$ & $1.6_{-0.6}^{+0.3}$ & $1.9\pm 0.2$ & $2.0_{-0.1}^{+0.2}$ & $1.14/39$ & {}\\
$15-06-2007$ & $1.2_{-0.4}^{+0.1}$ & $1.6_{-0.7}^{+0.5}$ & $1.8\pm 0.2$ & $2.0_{-0.1}^{+0.5}$ & $0.85/40$ & {}\\
$08-07-2007$ & $1.2\pm 0.1$   & $1.9\pm 0.4$ & $1.9\pm 0.2$ & $1.80\pm 0.03$ & $0.93/64$ & {}\\
$10-07-2007$ & $1.2_{-0.2}^{+0.1}$ & $1.9_{-0.4}^{+0.6}$ & $1.7_{-0.1}^{+0.2}$ & $1.9_{-0.1}^{+0.3}$ & $0.99/48$ & {}\\
$11-07-2007$ & $1.51\pm 0.06$ & {}      & {}         & $3.9\pm 0.3$ & $1.06/36$ & {}\\
$12-07-2007$ & $1.6\pm 0.1$ & $0.35_{-0.10}^{+0.08}$ & {} & $4.4_{-0.6}^{+0.7}$ & $0.94/37$ & {}\\
$13-07-2007$ & $1.2_{-0.2}^{+0.1}$ & $1.8_{-0.6}^{+1.0}$ & $1.8_{-0.2}^{+0.4}$ & $1.9\pm 0.1$ & $0.58/32$ & {}\\
$14-07-2007$ & $1.2_{-0.4}^{+0.1}$ & $1.5_{-0.8}^{+1.0}$ & $1.7_{-0.1}^{+0.3}$ & $2.2_{-0.1}^{+0.6}$ & $0.81/43$ & {}\\
\hline
\end{tabular}	
\normalsize
\label{tab:Summary1}
\end{table}

\begin{table}[!t]
\scriptsize
\caption{\scriptsize Summary of \emph{Swift}/XRT observations of ON~$+231$ ($z=0.102$, $N_{\rm H, Gal}=2.04\times 10^{20}$~cm$^{-2}$, Kalberla et al. 2005). Columns: (1) Date of observation [DD-MM-YYYY]; (2) Photon index (for the power-law model) or soft photon index (for the broken power-law model); (3) Break energy $E_{b}$ for the broken power-law model or folding energy $E_{f}$ of the exponential roll-off for the single power-law model plus low energy photon deficit [keV]; (4) Hard photon index for the broken power-law model; (5) Model normalization [$10^{-3}$~ph~cm$^{-2}$~s$^{-1}$~keV$^{-1}$]; (6) Reduced $\chi^2$ and degrees of freedom of the spectral fit; (7) Notes.}
\centering
\bigskip
\begin{tabular}{rrrrrrr}
\hline
Date       & $\Gamma$       & $E_{b}/E_{f}$ & $\Gamma_2$ & $N$  & $\tilde{\chi}^2$/dof & Notes\\
(1)         & (2)            & (3)           & (4)        & (5)  & (6)                  & (7)\\
\hline
$29-10-2005$ & $2.8\pm 0.1$   & $2.1_{-0.8}^{+1.2}$ & $1.3_{-4.3}^{+1.0}$ & $0.49\pm 0.04$ & $0.70/17$ & {}\\
$16-12-2005$ & $2.66\pm 0.09$ & {}            & {}         & $1.17\pm 0.06$ & $0.79/28$ & {}\\
$03-05-2007$ & $2.8\pm 0.3$   & {}            & {}         & $0.76\pm 0.10$ & $0.80/3$   & {}\\
$05-05-2007$ & $2.6\pm 0.1$   & {}            & {}         & $1.00\pm 0.08$ & $0.64/13$  & {}\\
$06-05-2007$ & $2.5\pm 0.1$   & {}            & {}         & $1.3\pm 0.1$   & $0.44/13$  & {}\\
$30-01-2008$ & $2.5_{-0.5}^{+0.4}$ & {}       & {}         & $0.39_{-0.7}^{+1.7}$ & $1.34/2$ & {}\\
$03-02-2008$ & $2.7\pm 0.4$   & {}            & {}         & $0.69\pm 0.13$ & $0.43/1$	& {}\\
$03-03-2008$ & $2.4\pm 0.2$   & {}            & {}         & $1.6_{-0.1}^{+0.3}$ & $1.53/8$ & {}\\
$14-03-2008$ & $2.8_{-0.2}^{+0.4}$ & $0.39_{-0.19}^{+0.13}$ & {} & $1.3\pm 0.3$ & $1.03/9$ & TeV det. (ATel 1422) \\
$16-03-2008$ & $2.4\pm 0.3$   & {}            & {}         & $0.82_{-0.10}^{+0.12}$ & $0.76/3$ & {}\\
$29-03-2008$ & $2.7\pm 0.4$   & {}            & {}         & $0.96\pm 0.15$ & $1.62/2$ & {}\\
$03-05-2008$ & $2.1\pm 0.3$   & {}            & {}         & $1.6\pm 0.2$ & $1.18/5$ & {}\\
\hline
\end{tabular}	
\normalsize
\label{tab:Summary2}
\end{table}

\begin{table}[!t]
\scriptsize
\caption{\scriptsize Summary of \emph{Swift}/XRT observations of S5~$0716+714$ ($z=0.31$, $N_{\rm H, Gal}=3.11\times 10^{20}$~cm$^{-2}$, Kalberla et al. 2005). Columns: (1) Date of observation [DD-MM-YYYY]; (2) Photon index (for the power-law model) or soft photon index (for the broken power-law model); (3) Break energy $E_{b}$ for the broken power-law model or folding energy $E_{f}$ of the exponential roll-off for the single power-law model plus low energy photon deficit [keV]; (4) Hard photon index for the broken power-law model; (5) Model normalization [$10^{-3}$~ph~cm$^{-2}$~s$^{-1}$~keV$^{-1}$]; (6) Reduced $\chi^2$ and degrees of freedom of the spectral fit; (7) Notes.}
\centering
\bigskip
\begin{tabular}{rrrrrrr}
\hline
Date        & $\Gamma$       & $E_{b}/E_{f}$ & $\Gamma_2$ & $N$  & $\tilde{\chi}^2$/dof & Notes\\
(1)         & (2)            & (3)           & (4)        & (5)  & (6)                  & (7)\\
\hline
$02-04-2005$ & $2.61\pm 0.08$ & {}            & {}         & $1.44\pm 0.08$ & $1.19/36$ & {}\\
$18-08-2005$ & $2.38\pm 0.08$ & $2.5_{-1.4}^{+1.1}$ & $1.8_{-0.5}^{+0.4}$ & $2.60_{-0.08}^{+0.10}$ & $1.03/59$ & {}\\
$23-10-2007$ & $2.28\pm 0.09$ & {}            & {}         & $3.1\pm 0.2$ & $0.96/25$ & Start of AGILE Campaign\\
$24-10-2007$ & $2.2\pm 0.1$   & {}            & {}         & $1.9\pm 0.2$ & $0.58/9$ & {}\\
$25-10-2007$ & $2.1\pm 0.1$   & {}            & {}         & $2.5\pm 0.1$ & $0.51/22$ & {}\\
$26-10-2007$ & $2.4\pm 0.3$   & $1.1_{-0.2}^{+0.8}$ & $1.8_{-0.3}^{+0.2}$ & $2.0\pm 0.3$ & {}\\
$27-10-2007$ & $2.43\pm 0.08$ & {}            & {}         & $5.2\pm 0.3$ & $0.90/34$ & {}\\
$28-10-2007$ & $2.2\pm 0.1$   & {}            & {}         & $2.1\pm 0.2$ & $0.62/11$ & {}\\
$03-11-2007$ & $2.44\pm 0.06$ & {}            & {}         & $6.0\pm 0.2$ & $1.09/53$ & {}\\
$06-11-2007$ & $2.3\pm 0.1$   & {}            & {}         & $1.8\pm 0.1$ & $0.57/11$ & {}\\
$09-11-2007$ & $2.38_{-0.09}^{+0.17}$ & {}    & {}         & $2.5_{-0.1}^{+0.3}$ & $1.24/17$ & {}\\
$13-11-2007$ & $2.4\pm 0.1$   & {}            & {}         & $1.8\pm 0.1$ & $0.89/14$ & End of AGILE Campaign\\
$28-04-2008$ & $2.1_{-0.1}^{+0.3}$ & $0.70_{-0.14}^{+0.25}$ & $2.6\pm 0.1$ & $11_{-2}^{+3}$ & $1.09/52$ & ATel 1495, 1500 (TeV det.)\\
$29-04-2008$ & $2.49\pm 0.05$ & $2.6_{-0.3}^{+0.5}$ & $3.4_{-0.4}^{0.6}$ & $7.9\pm 0.2$ & $1.16/93$ & ATel 1500 (TeV det.)\\
$02-05-2008$ & $2.5\pm 0.1$  & {}             & {}         & $2.9\pm 0.2$ & $0.89/17$ & {} \\
\hline
\end{tabular}	
\normalsize
\label{tab:Summary3}
\end{table}

\begin{table}[!t]
\scriptsize
\caption{\scriptsize Summary of \emph{Swift}/XRT observations of PKS~$2155-304$ ($z=0.116$, $N_{\rm H, Gal}=1.48\times 10^{20}$~cm$^{-2}$, Kalberla et al. 2005). Columns: (1) Date of observation [DD-MM-YYYY]; (2) Photon index (for the power-law model) or soft photon index (for the broken power-law model); (3) Break energy $E_{b}$ for the broken power-law model or folding energy $E_{f}$ of the exponential roll-off for the single power-law model plus low energy photon deficit [keV]; (4) Hard photon index for the broken power-law model; (5) Model normalization [$10^{-2}$~ph~cm$^{-2}$~s$^{-1}$~keV$^{-1}$]; (6) Reduced $\chi^2$ and degrees of freedom of the spectral fit; (7) Notes.}
\centering
\bigskip
\begin{tabular}{rrrrrrr}
\hline
Date        & $\Gamma$       & $E_{b}/E_{f}$ & $\Gamma_2$ & $N$  & $\tilde{\chi}^2$/dof & Notes\\
(1)         & (2)            & (3)           & (4)        & (5)  & (6)                  & (7)\\
\hline
$17-11-2005$ & $2.6\pm 0.1$  & {}            & {}         & $2.9\pm 0.2$ & $0.78/21$ & {}\\
$11-04-2006$ & $2.2\pm 0.3$  & {}            & {}         & $2.2\pm 0.3$ & $0.96/3$ & {}\\
$16-04-2006$ & $2.48\pm 0.08$ & {}           & {}         & $2.8\pm 0.1$ & $1.14/36$ & {}\\
$20-04-2006$ & $2.4\pm 0.1$   & {}           & {}         & $3.8\pm 0.2$ & $1.04/19$ & {}\\
$30-04-2006$ & $2.38\pm 0.03$ & {}           & {}         & $1.56\pm 0.04$ & $1.05/122$ & {}\\
$29-07-2006$ & $2.44\pm 0.02$ & $1.5_{-0.1}^{+0.3}$ & $2.76_{-0.04}^{+0.09}$ & $8.36_{-0.11}^{+0.08}$ & $1.09/312$ & Giant TeV Flare\\
$01-08-2006$ & $2.59\pm 0.07$ & $1.7_{-0.5}^{+1.7}$ & $3.0_{-0.2}^{+1.2}$ & $6.8\pm 0.2$ & $0.92/86$ & {}\\
$02-08-2006$ & $2.58\pm 0.03$ & $2.0\pm 0.4$ & $3.0_{-0.1}^{+0.2}$ & $5.91\pm 0.09$ & $1.14/191$ & {}\\
$03-08-2006$ & $2.42\pm 0.04$ & $1.4_{-0.3}^{+0.4}$ & $2.72_{-0.07}^{+0.11}$ & $7.1\pm 0.1$ & $1.09/206$ & {}\\
$05-08-2006$ & $2.74\pm 0.05$ & {}           & {}         & $6.0\pm 0.2$ & $1.18/87$ & {}\\
$06-08-2006$ & $2.74\pm 0.07$ & {}           & {}         & $5.2\pm 0.2$ & $0.99/46$ & {}\\
$10-08-2006$ & $2.60\pm 0.06$ & {}           & {}         & $6.4\pm 0.2$ & $0.80/61$ & {}\\
$11-08-2006$ & $2.46\pm 0.08$ & {}           & {}         & $5.7\pm 0.3$ & $0.92/27$ & {}\\
$12-08-2006$ & $2.8\pm 0.1$   & {}           & {}         & $3.5\pm 0.3$ & $0.96/15$ & {}\\
$14-08-2006$ & $2.61\pm 0.09$ & {}           & {}         & $3.1\pm 0.2$ & $1.37/29$ & {}\\
$15-08-2006$ & $2.6\pm 0.1$   & {}           & {}         & $2.5\pm 0.1$ & $0.37/20$ & {}\\
$17-08-2006$ & $2.6\pm 0.1$   & {}           & {}         & $2.3\pm 0.1$ & $0.71/22$ & {}\\
$18-08-2006$ & $2.62\pm 0.08$ & {}           & {}         & $2.8\pm 0.1$ & $1.27/34$ & {}\\
$19-08-2006$ & $2.5\pm 0.2$   & {}           & {}         & $2.7\pm 0.2$ & $1.12/10$ & {}\\
$20-08-2006$ & $2.6\pm 0.1$   & {}           & {}         & $3.5\pm 0.2$ & $1.10/19$ & {}\\
$21-08-2006$ & $2.5\pm 0.3$   & {}           & {}         & $2.8\pm 0.3$ & $0.23/4$  & {}\\
$22-08-2006$ & $2.7\pm 0.1$   & {}           & {}         & $4.6\pm 0.3$ & $1.18/20$ & {}\\
$23-08-2006$ & $2.5\pm 0.2$   & {}           & {}         & $2.6\pm 0.2$ & $1.82/10$ & {}\\
$26-08-2006$ & $2.6\pm 0.1$   & {}           & {}         & $1.7\pm 0.1$ & $1.64/16$ & {}\\
$27-08-2006$ & $2.7\pm 0.1$   & {}           & {}         & $1.8\pm 0.1$ & $0.73/19$ & {}\\
$28-08-2006$ & $2.3\pm 0.4$   & {}           & {}         & $3.6\pm 0.5$ & $1.02/3$ & {}\\
$29-08-2006$ & $2.5\pm 0.4$   & {}           & {}         & $3.2\pm 0.5$ & $0.68/2$ & {}\\
$22-04-2007$ & $2.64\pm 0.04$ & {}           & {}         & $3.9\pm 0.1$ & $0.75/90$ & {}\\
$12-05-2008$ & $2.14\pm 0.03$ & $2.3\pm 0.3$ & $2.9\pm 0.2$ & $3.77\pm 0.07$ & $1.00/180$ & {}\\
$12-05-2008$ & $2.53\pm 0.02$ & {}           & {}         & $4.40\pm 0.05$ & $1.10/229$ & {}\\
$13-05-2008$ & $2.08\pm 0.04$ & $1.9_{-0.2}^{+0.3}$ & $2.7\pm 0.1$ & $4.36\pm 0.08$ & $0.92/201$ & {}\\
\hline
\end{tabular}	
\normalsize
\label{tab:Summary4}
\end{table}

Optical and ultraviolet data of UVOT (Roming et al. 2005) in the filters $B$ ($4329$~\AA) and $UVW1$ ($2634$~\AA) were considered. The snapshots of every individual pointing were integrated with \texttt{uvotimsum} task and then analyzed with the \texttt{uvotsource} task. The source region was a circle with radius of $5''$ for $B$ filter and $10''$ for $UVW1$ filter. The background was extracted from an annular region centered on the blazar, with internal radius of $7''$ for optical filters and $12''$ for UV. The external radius was selected on the basis of the presence or not of nearby contaminating sources and was $40''$ for ON~$+231$, $50''$ for 3C~$279$ and PKS~$2155-304$, $60''$ for S5~$0716+714$. 

\begin{figure}
\begin{center}
\includegraphics[angle=270,scale=0.4]{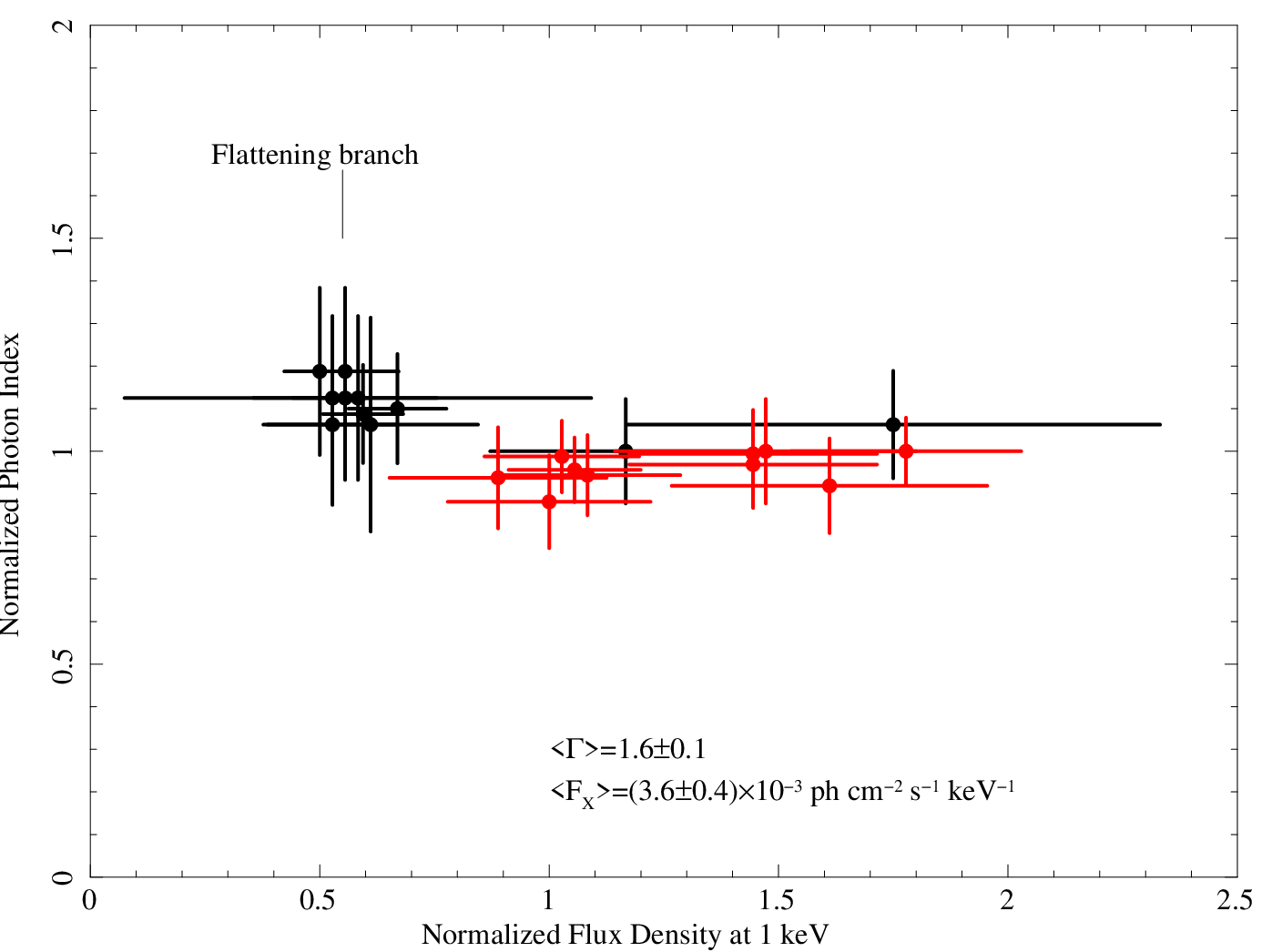}\\
\includegraphics[angle=270,scale=0.4]{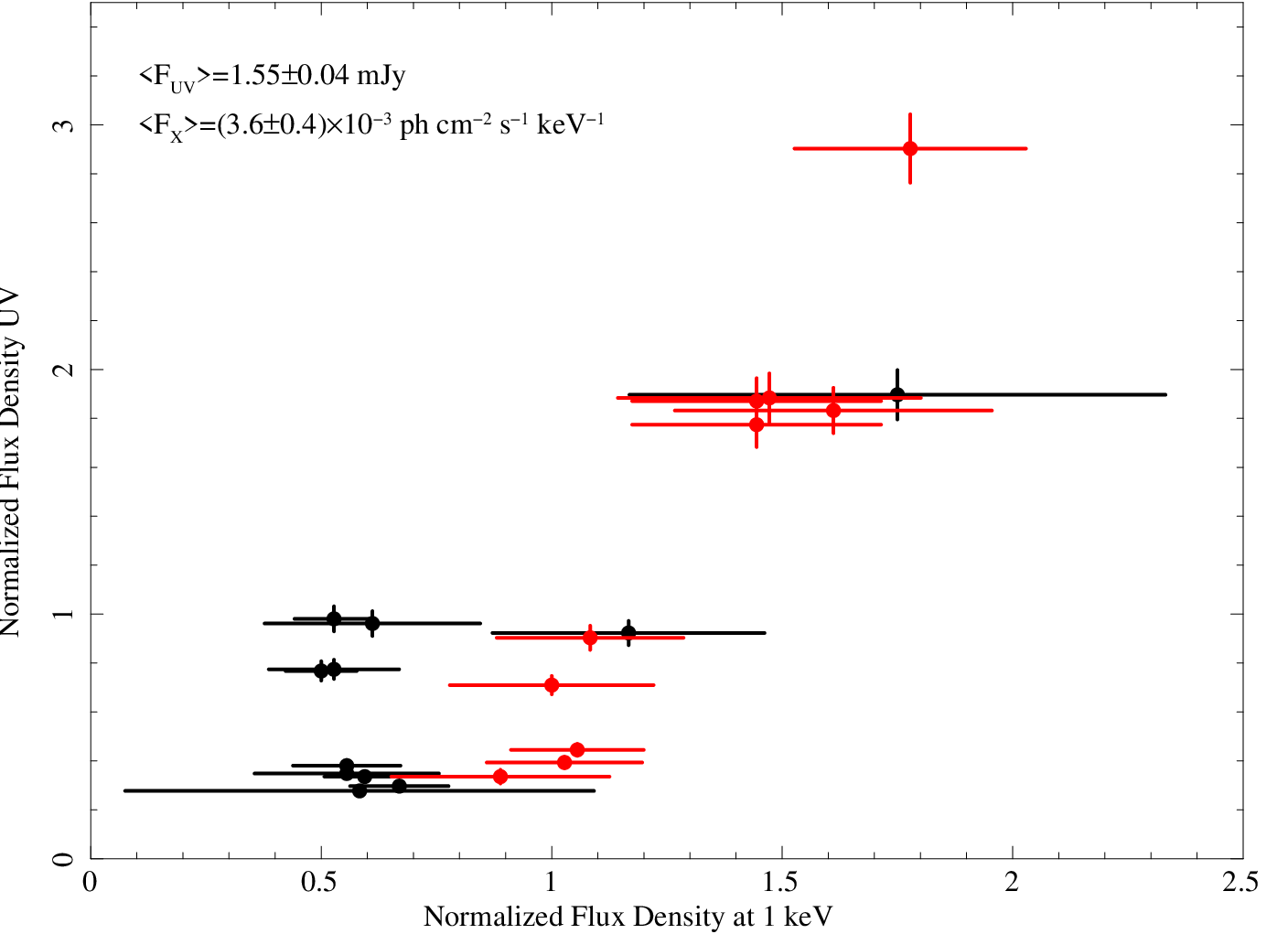}\\
\includegraphics[angle=270,scale=0.4]{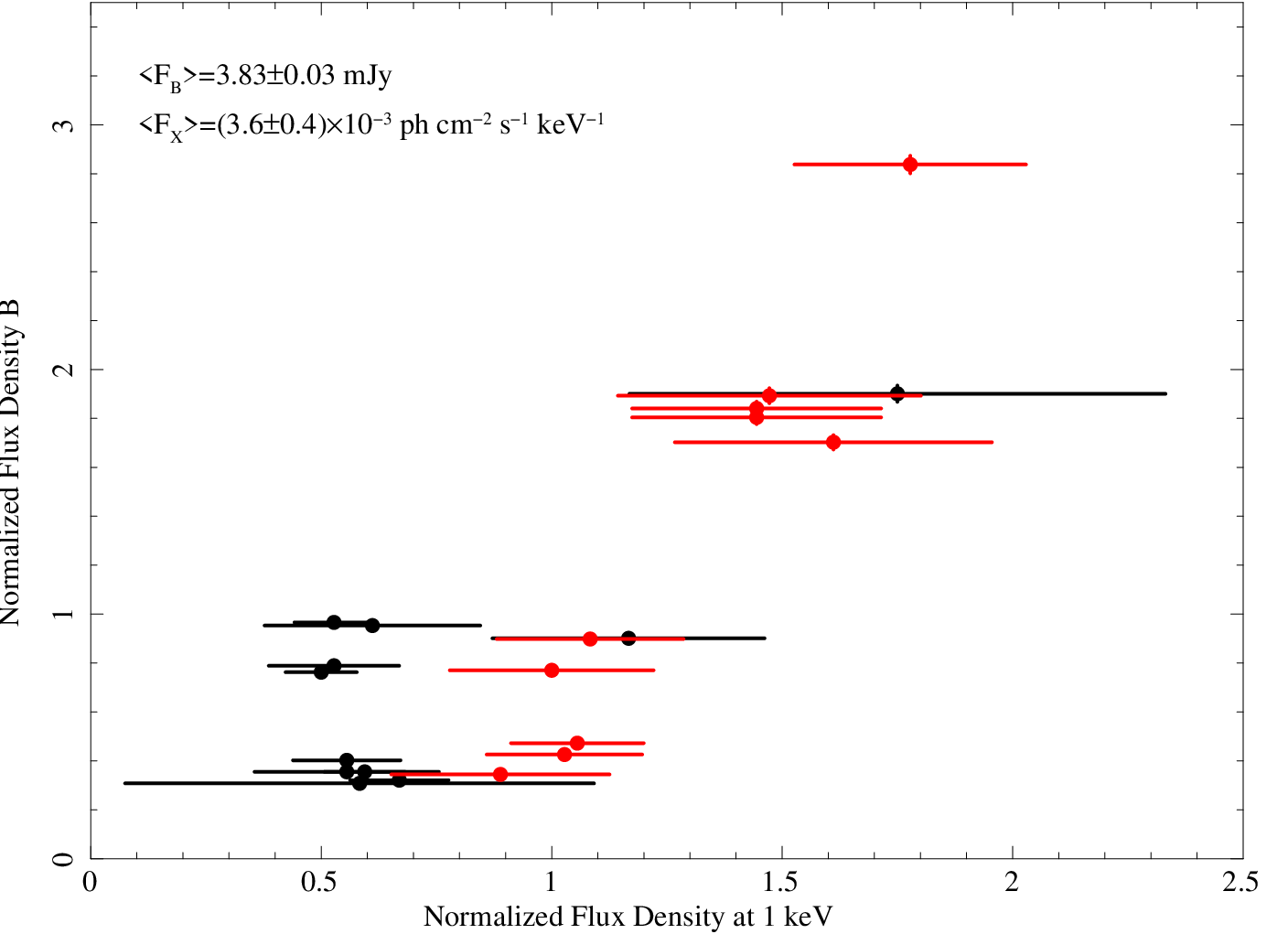}\\
\end{center}
\caption{\scriptsize X-ray flux density at $1$~keV versus $\Gamma$ (\emph{top panel}), UV flux density at $2634$~\AA (\emph{center panel}), and optical flux density at $4329$~\AA (\emph{bottom panel}) for 3C~$279$.}
\label{fig:3C279}
\end{figure}

\begin{figure}
\begin{center}
\includegraphics[angle=270,scale=0.4]{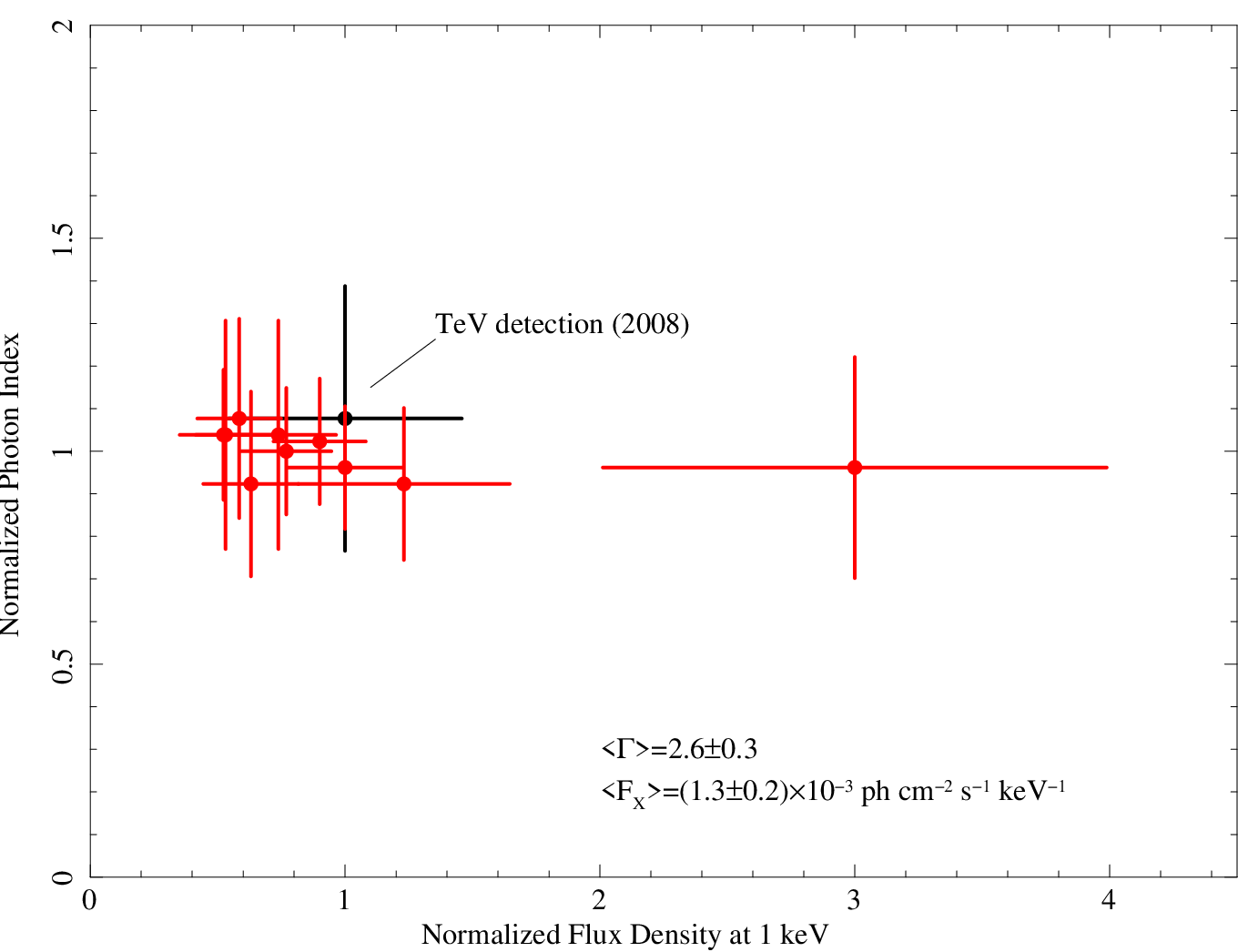}\\
\includegraphics[angle=270,scale=0.4]{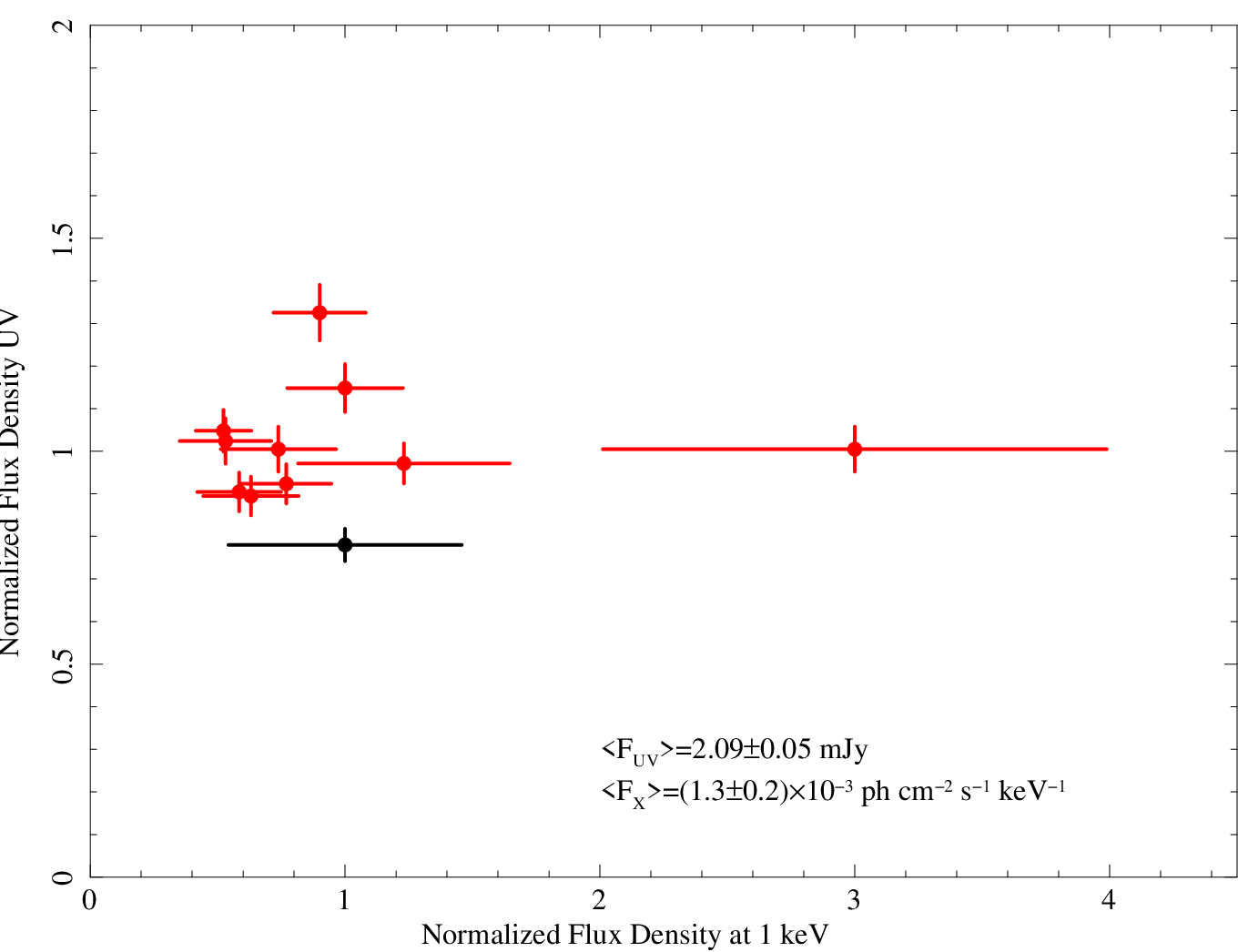}\\
\includegraphics[angle=270,scale=0.4]{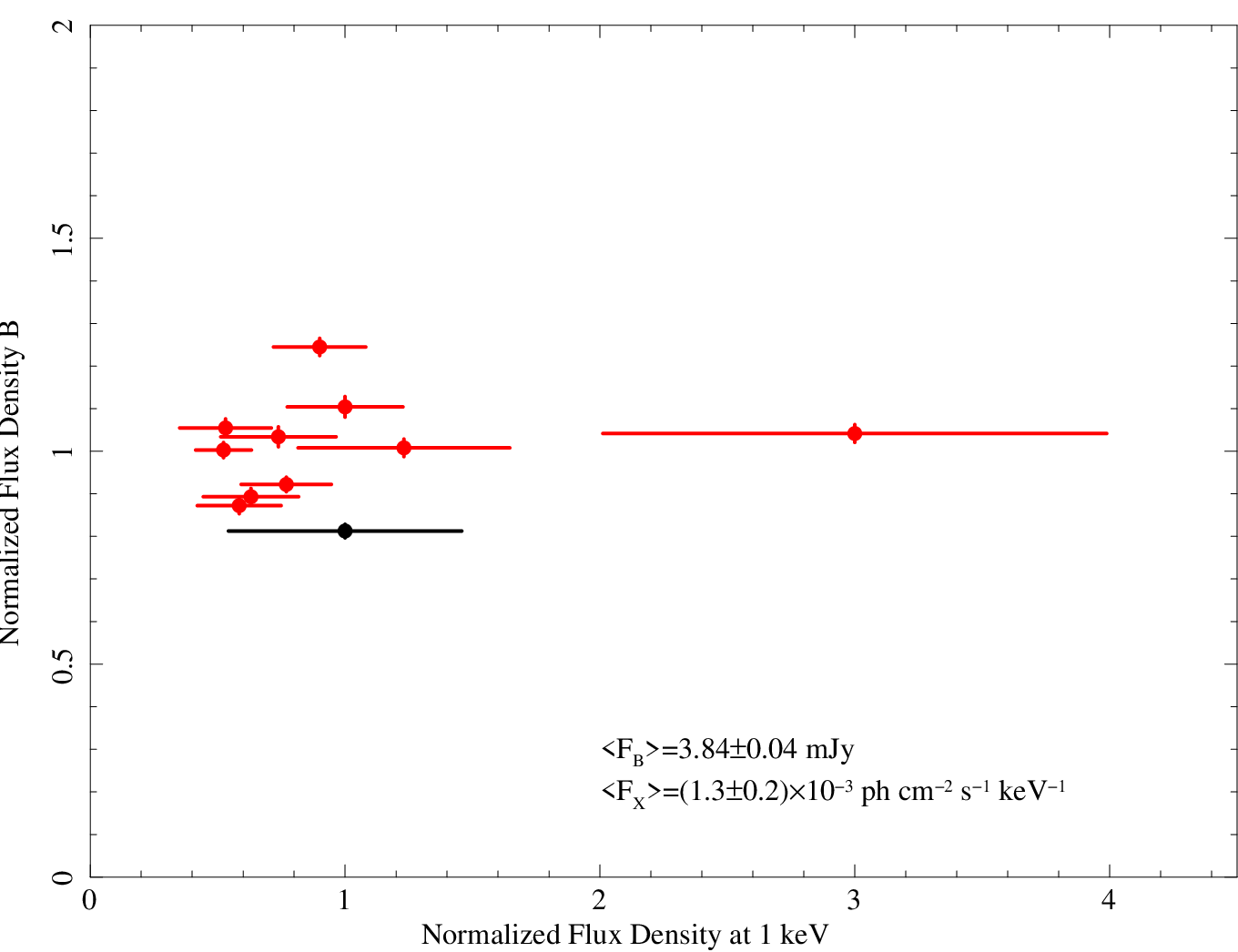}\\
\end{center}
\caption{\scriptsize X-ray flux density at $1$~keV versus $\Gamma$ (\emph{top panel}), UV flux density at $2634$~\AA (\emph{center panel}), and optical flux density at $4329$~\AA (\emph{bottom panel}) for ON~$+231$.}
\label{fig:ON231}
\end{figure}

\begin{figure}
\begin{center}
\includegraphics[angle=270,scale=0.4]{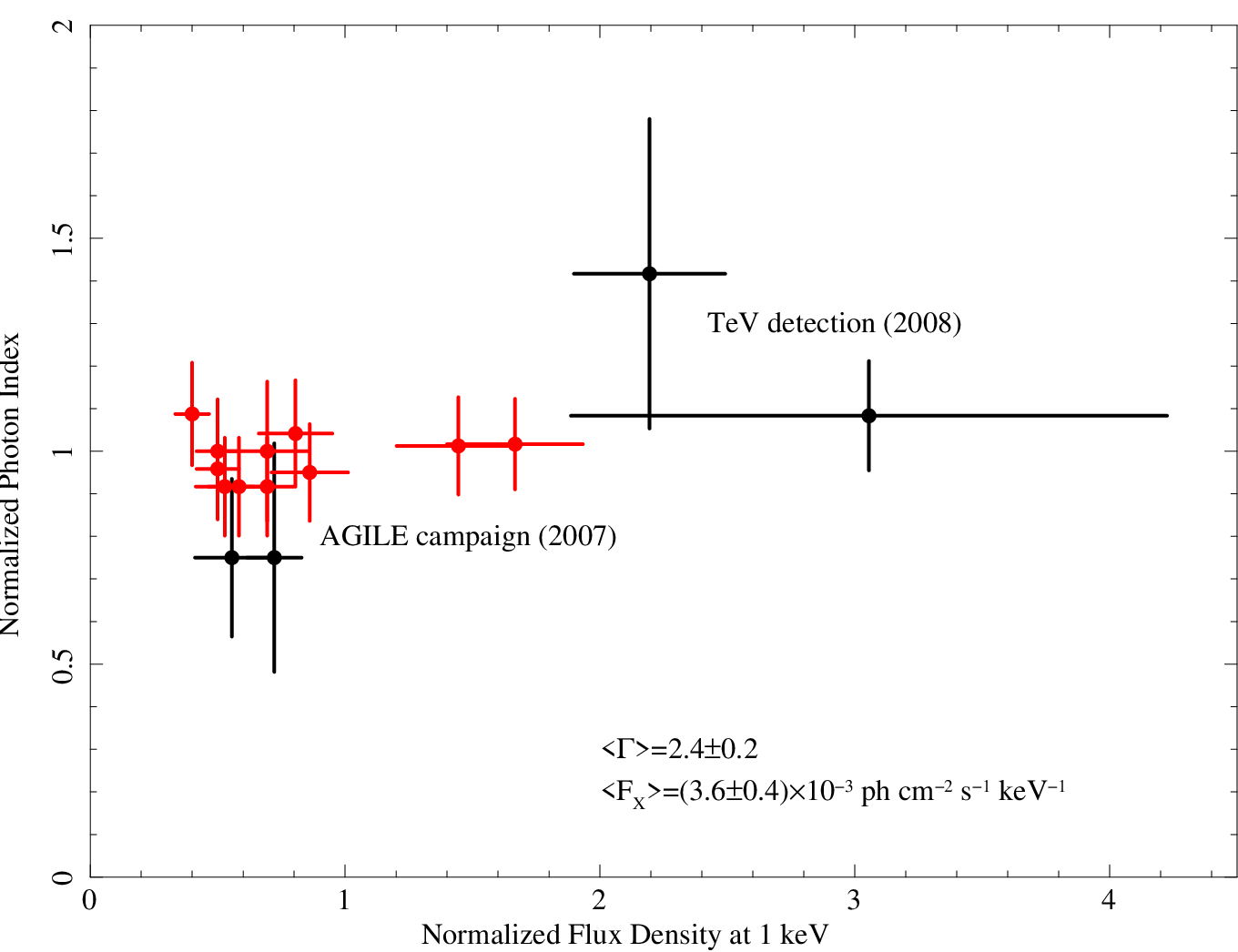}\\
\includegraphics[angle=270,scale=0.4]{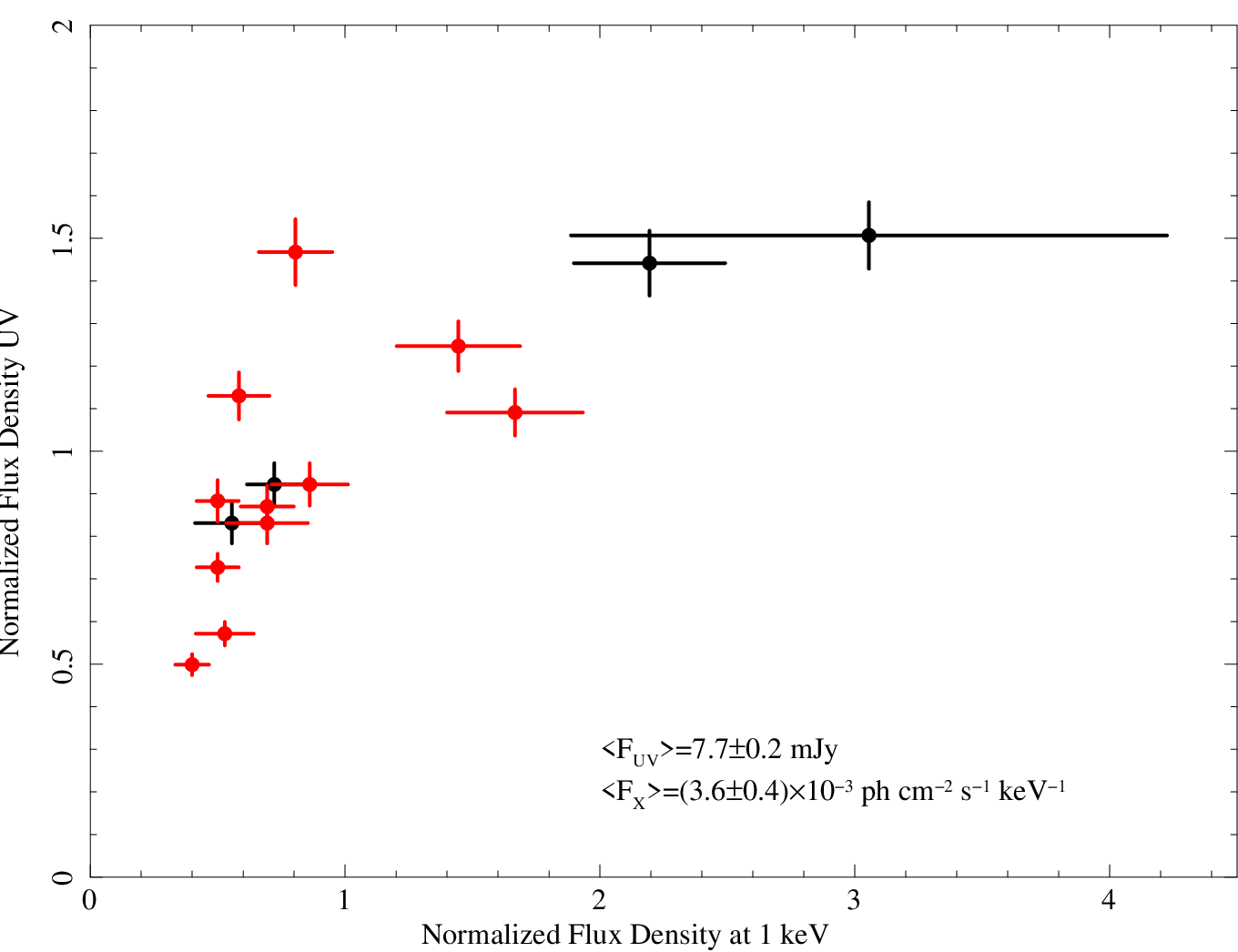}\\
\includegraphics[angle=270,scale=0.4]{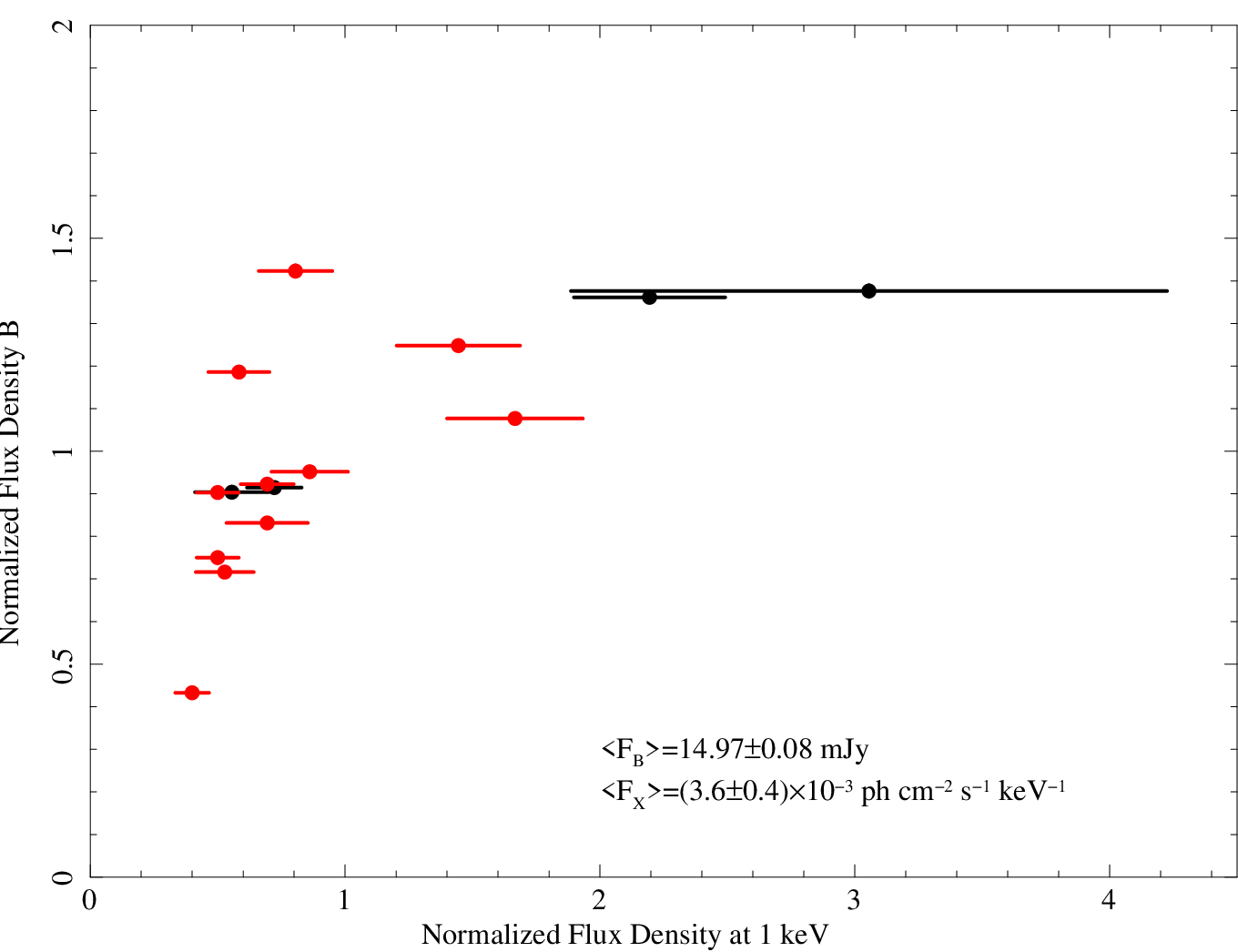}\\
\end{center}
\caption{\scriptsize X-ray flux density at $1$~keV versus $\Gamma$ (\emph{top panel}), UV flux density at $2634$~\AA (\emph{center panel}), and optical flux density at $4329$~\AA (\emph{bottom panel}) for S5~$0716+71$.}
\label{fig:S50716}
\end{figure}

\begin{figure}
\begin{center}
\includegraphics[angle=270,scale=0.4]{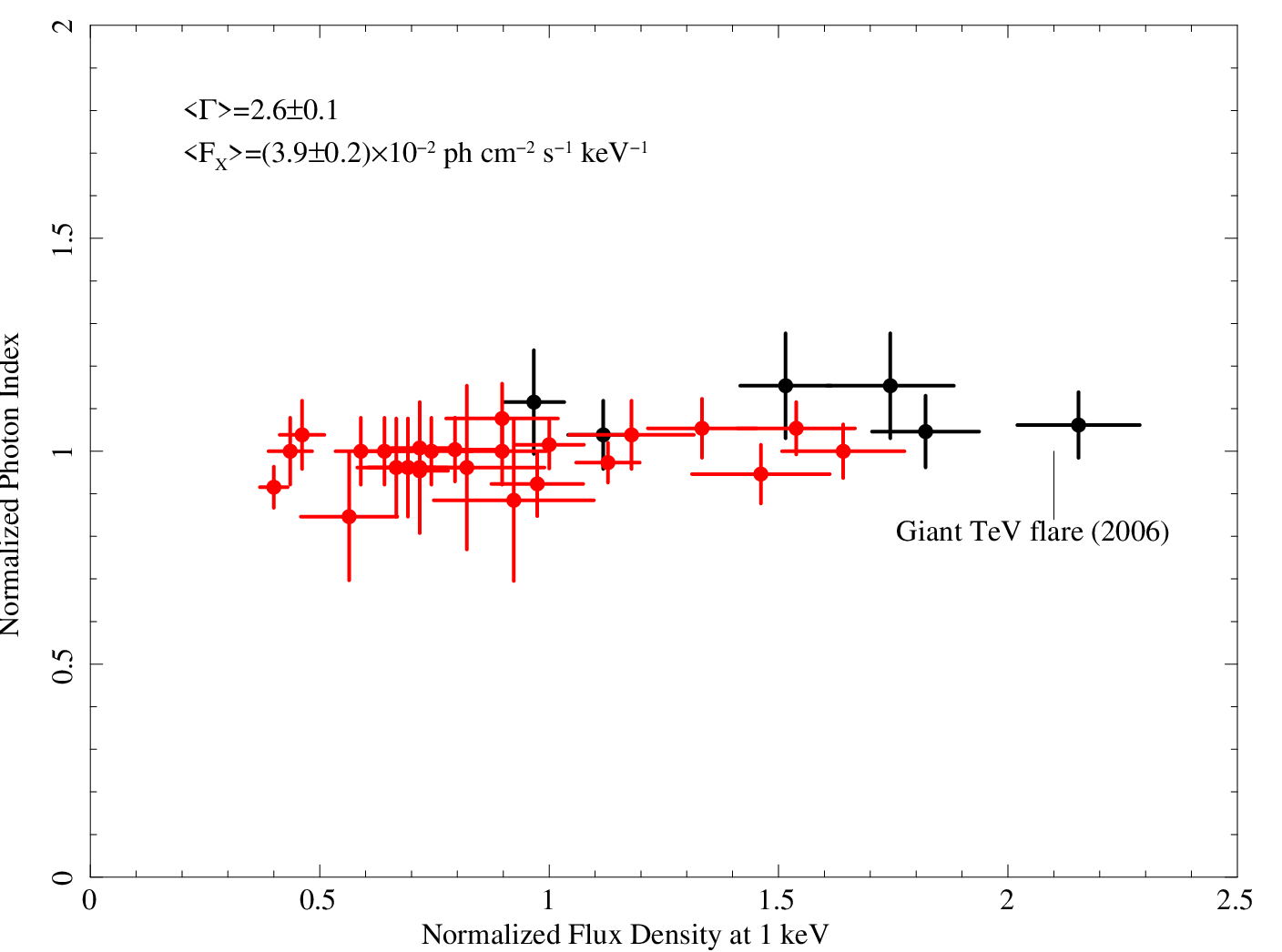}\\
\includegraphics[angle=270,scale=0.4]{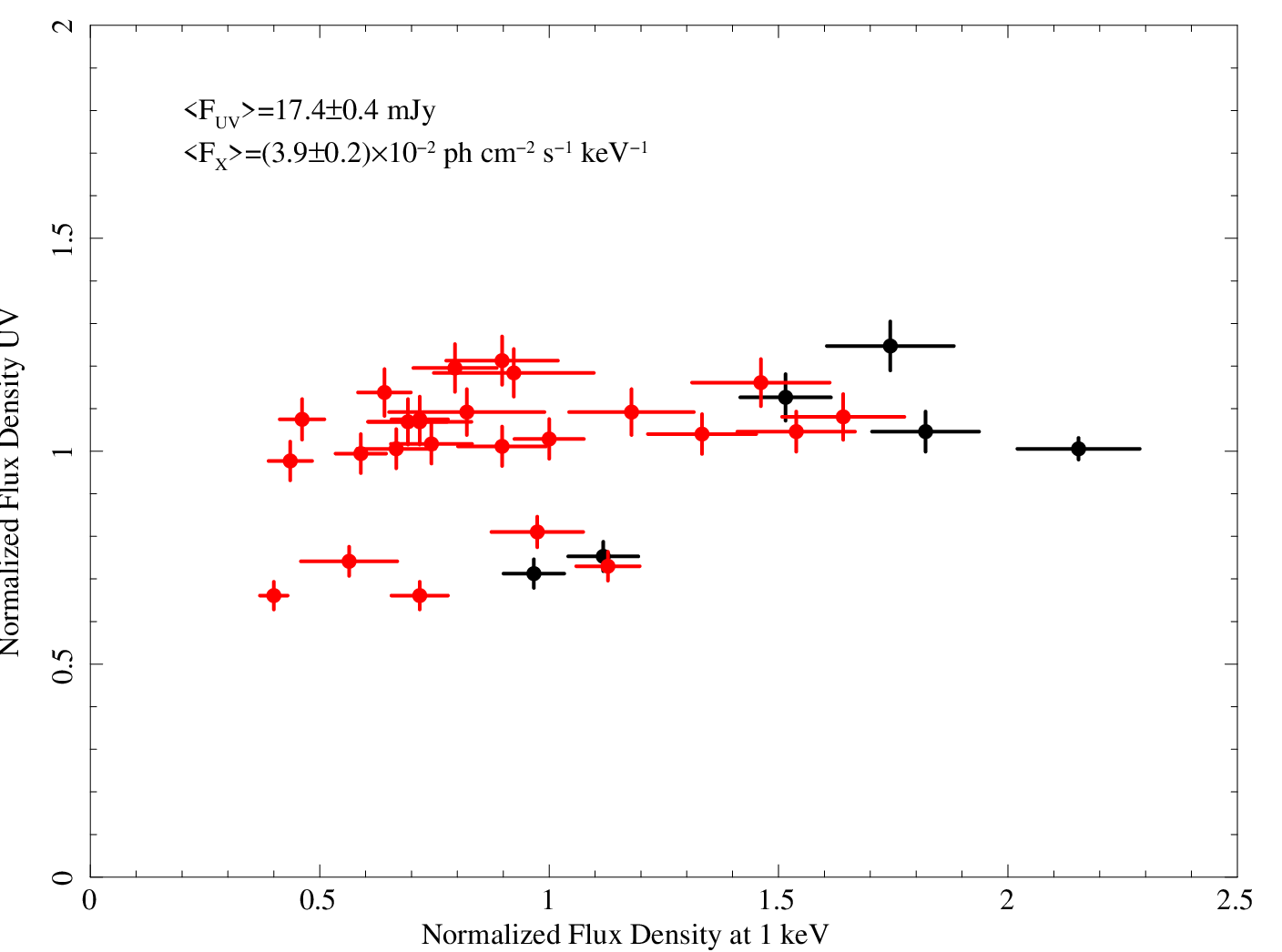}\\
\includegraphics[angle=270,scale=0.4]{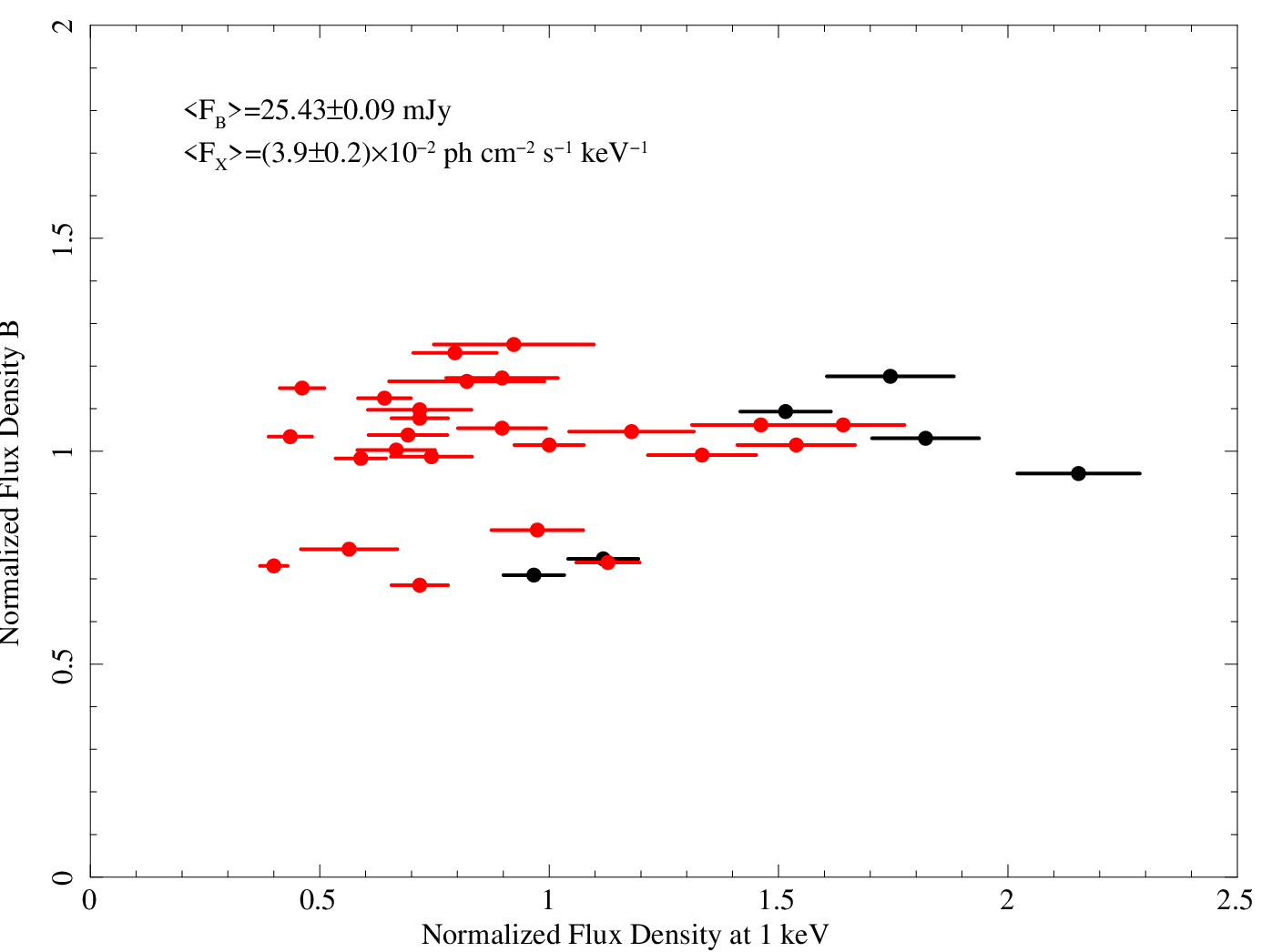}\\
\end{center}
\caption{\scriptsize X-ray flux density at $1$~keV versus $\Gamma$ (\emph{top panel}), UV flux density at $2634$~\AA (\emph{center panel}), and optical flux density at $4329$~\AA (\emph{bottom panel}) for PKS~$2155-304$.}
\label{fig:PKS2155}
\end{figure}

To emphasize the source variability and minimize the systematic or contaminating effects, the selected parameters (X-ray, UV and Optical flux densities, $\Gamma$) have been normalized to their average value (indicated between $<>$). The results are shown in Fig.~\ref{fig:3C279}, \ref{fig:ON231}, \ref{fig:S50716} and \ref{fig:PKS2155}, where red points refer to X-ray spectra fitted with a single power-law model and black points indicate data fitted with a broken power-law model or with a single power-law with low-energy exponential roll-off. In the case of a broken power-law model, the photon index used in the figure is the soft one, i.e. below the energy break.

\section{Discussion}
As expected, all the sources display correlated variability from optical to X-ray frequencies, since the radiation emitted by a blazar is dominated by the relativistic jet component, but there are some specific points worth noting.

The most interesting case is 3C $279$, whose observations were performed during two MW campaigns, one in mid-January 2006 (B\"ottcher et al. 2007) triggered by high optical state ($R = 14.5$), while the other was performed in January 2007, with an even higher optical state ($R\sim 13$). The black points in Fig.~\ref{fig:3C279}, \emph{left column}, indicate the deficit of photons in the low energy part of the X-ray spectrum, which appears to be confined in a ``branch'' separated by the remaining. This photon deficit can be modeled either with a broken power-law with $\Gamma_{soft}<\Gamma_{hard}$ (low-energy spectral flattening) and a break between $\sim 1.5-1.9$~keV, or with a single power-law with low-energy exponential roll-off and folding energy of $\sim 0.35$~keV. These changes occur on timescale of days. A similar effect, i.e. changes in the low-energy spectral flattening, has been already noted by Bianchin et al. (2008) in the case of the blazar PKS~$2149-306$ ($z=2.345$). 

The nature of this flattening, often observed in high-redshift blazars, has been examined and is still under investigation. The low-energy spectrum ($E\lesssim 2$~keV) can give important information about the very low end of the electron distribution, since -- according to the EC models -- a break is expected at $\nu_{\rm br,obs} \simeq \nu_{\rm BB}\Gamma_{\rm bulk}^{2}\gamma_{\rm min}/(1+z)$, where $\Gamma_{\rm bulk}$ is the bulk Lorentz factor, $\gamma_{\rm min}$ is the minimum Lorentz factor of the electron distribution, $\nu_{\rm BB}$ is the peak frequency of the emission of the accretion disk (modeled as a black body). In the case of 3C~$279$ and many other flat-spectrum radio quasars (FSRQ), the typical values of the parameters are (e.g. Ballo et al. 2002): $\Gamma_{\rm bulk}=10-20$, $\gamma_{\rm min}=1$, $\nu_{\rm BB}\sim 10^{15}$~Hz, $z=0.5362$, thus resulting in $\nu_{\rm br,obs}\sim 10^{17}$~Hz. A more detailed discussion of the low-energy photon deficit in the framework of a natural curvature of the blazar continuum can be found in Tavecchio et al. (2007) and Ghisellini et al. (2007). 

In reality, things are not so easy and self-explaining: indeed, at these energies there could be contributions from SSC (as in the case of 3C~$279$, Ballo et al. 2002) or from other processes, like the bulk motion Comptonization (Celotti et al. 2007). In addition, since this break appears in the X-ray spectrum as a photon deficit at low-energies, some authors suggested that it can be explained as an absorption (cold or ionized) in addition to the Galactic column and intrinsic to the source (e.g. Cappi et al. 1997, Fiore et al. 1998, Piconcelli \& Guainazzi 2005, Page et al. 2005, Yuan et al. 2006). 

The data on 3C~$279$ shown in Fig.~\ref{fig:3C279}, \emph{left column}, favor the hypothesis of the natural break in the SED, which is the curvature of the continuum of a typical FSRQ. The points of the flattening branch (Fig.~\ref{fig:3C279}) are correlated each other at all the analyzed frequencies (optical to X-rays), as expected from the emission of the relativistic jet. Would it be something external to the jet, like an absorption intrinsic to the source, it is expected to find no correlation with the changes in the continuum of the source. 

Also the black point in ON~$+231$ (Fig.~\ref{fig:3C279}, \emph{right column}) indicates a flattening at low energies in the X-ray spectrum and refers to an observation performed on March $14$, $2008$, just one day after the detection of a VHE flare by \emph{VERITAS} (Swordy et al. 2008). However, since ON~$+231$ is an intermediate blazar, it is not expected to have the same behavior of 3C~$279$. One single point is not sufficient for a fruitful investigation, but surely the fact that the low-energy flattening occurred in proximity of a VHE flare, make it an interesting case worth studying.

On the other hand, the black points in S5~$0716+71$ (Fig.~\ref{fig:PKS2155}, \emph{left column}) indicate fit with a broken power-law model, with $\Gamma_{soft}>\Gamma_{hard}$, where the soft component is due to the tail of the synchrotron emission and the hard X-ray emission is due to the emergence of the inverse-Compton component (e.g. Foschini et al. 2006). The points at highest X-ray flux were measured a few days after the VHE detection by \emph{MAGIC} (Teshima et al. 2008), while the observations during the \emph{AGILE} pointings had lower X-ray fluxes (cf. Giommi et al. 2008). 

Interestingly, despite the fact that both ON~$+231$ and S5~$0716+71$ are intermediate blazars, the former was detected at TeV energies during a low optical/UV state, while the contrary occurred for the latter. 

PKS~$2155-304$ was extensively observed during the giant TeV flares in July 2006 and a more detailed analysis has been presented elsewhere (Foschini et al. 2007). Here I confirm the previous analyses.

\section{Conclusion}
I started a systematic study of the simultaneous optical-to-X-ray variability of blazars and presented here the preliminary results. The most interesting result concerns the FSRQ 3C~$239$, which shows a clear pattern when a low-energy photon deficit appears in the X-ray spectrum. Although this deficit has been interpreted in different ways, the data available here favor the hypothesis that is due to the natural curvature of the continuum in a typical FSRQ. 

Future steps of this research are to enlarge the sample and continue the monitoring, taking advantage of the unique capabilities of the \emph{Swift} satellite. In addition, the successful beginning of the operations of the \emph{Fermi Gamma-ray Space Telescope} during the summer of 2008, will give, in the near future, the possibility to include also the data at $\gamma-$rays, which are of paramount importance in defining the true state of activity of the source.

\textbf{References}\\
\scriptsize
Ballo, L., Maraschi, L., Tavecchio, F., et al., Spectral energy distributions of 3C 279 revisited: BeppoSAX observations and variability models, ApJ, 567, 50-57, 2002.\\
Bianchin, V., Foschini, L., Ghisellini, G., et al., The changing look of PKS~$2149-306$, A\&A, accepted for publication.\\
B\"ottcher, M., Basu, S., Joshi, M., et al., The WEBT campaign on the blazar 3C~$279$ in $2006$, ApJ, 670, 968-977, 2007.\\
Burrows, D.N., Hill, J.E., Nousek, J.A., et al., The Swift X-Ray Telescope, Space Sci. Rev., 120, 165-195, 2005.\\
Cappi, M., Matsuoka, M., Comastri, A., et al., ASCA and ROSAT X-Ray Spectra of High-Redshift Radio-loud Quasars, ApJ, 478, 492-510, 1997.\\
Celotti, A., Ghisellini, G., Fabian, A.C., Bulk Comptonization spectra in blazars, MNRAS, 375, 417-424, 2007.\\
Fiore F., Elvis M., Giommi P. \& Padovani P., X-Ray Spectral Survey of WGACAT Quasars. I. Spectral Evolution and Low-Energy Cutoffs, ApJ, 492, 79-90, 1998.\\
Foschini, L., Tagliaferri, G., Pian, E., et al., Simultaneous X-ray and optical observations of S5~$0716+714$ after the outburst of March~$2004$, A\&A, 455, 871-877, 2006.\\
Foschini, L., Ghisellini, G., Tavecchio, F., et al., X-ray/UV/Optical follow-up of the blazar PKS~$2155-304$ after the giant TeV flares of July~$2006$, ApJ, 657, L81-L84, 2007.\\
Fossati, G., Maraschi, L., Celotti, A., Comastri, A., \& Ghisellini, G., A unifying view of the spectral energy distribution of blazars, MNRAS, 299, 433-448, 1998.\\
Gehrels, N., Chincarini, G., Giommi, P., et al., The Swift Gamma-Ray Burst Mission, ApJ, 611, 1005-1020, 2004.\\
Ghisellini, G., Celotti, A., Fossati, G., Maraschi, L., \& Comastri, A., A theoretical unifynig scheme for gamma-ray bright blazars, MNRAS, 301, 451-468, 1998.\\
Ghisellini, G., Foschini, L., Tavecchio, F. \& Pian, E., On the 2007 July flare of the blazar 3C 454.3, MNRAS, 382, L82-L86, 2007.\\
Ghisellini, G. \& Tavecchio, F., The blazar sequence: a new perspective, MNRAS, 387, 1669-1680, 2008.\\
Giommi, P., Colafrancesco, S., Cutini, S., et al., AGILE and Swift simultaneous observations of the blazar S5~$0716+714$ during the bright flare of October~$2007$, A\&A, 487, L49-L52, 2008.\\
Kalberla, P. M. W., Burton, W. B., Hartmann, Dap, et al., The Leiden/Argentine/Bonn (LAB) Survey of Galactic HI. Final data release of the combined LDS and IAR surveys with improved stray-radiation corrections, A\&A, 440, 775-782, 2005.\\
Maraschi, L., Foschini, L., Ghisellini, G., Tavecchio, F., Sambruna, R.M., Testing the blazar spectral sequence: X-ray selected blazars, MNRAS, accepted for publication, 2008, [\texttt{arXiv:0810.0145}].\\
Padovani, P., The blazar sequence: validity and predictions, Astrophys. Space Sci., 309, 63-71, 2007.\\
Page, K. L., Reeves, J. N., O'Brien, P. T. \& Turner, M. J. L., XMM-Newton spectroscopy of high-redshift quasars, MNRAS, 364, 195-207, 2005.\\
Piconcelli, E. \& Guainazzi, M., XMM-Newton discovery of soft X-ray absorption in the high-z superluminous blazar RBS 315, A\&A, 442, L53-L56, 2005.\\
Roming, P.W.A., Kennedy, T.E., Mason, K.O., et al., The Swift Ultra-Violet/Optical Telescope, Space Sci. Rev., 120, 95-142, 2005.\\
Swordy, S. for the VERITAS Collaboration, VERITAS discovers TeV gamma rays from W Comae, ATel 1422, March 14, 2008.\\
Tavecchio, F., Maraschi, L., Ghisellini, G., et al., Low-energy cutoff and hard X-ray spectra in high-z radio-loud quasars: the Suzaku view of RBS315, ApJ, 665, 980-989, 2007.\\
Teshima, M. for the MAGIC Collaboration, MAGIC discovers VHE gamma ray emission from the blazar S5~$0716+714$, ATel 1500, April 30, 2008.\\
Yuan, W., Fabian, A. C., Worsley, M. A. \& McMahon, R. G., X-ray spectral properties of high-redshift radio-loud quasars beyond redshift 4 - first results, MNRAS, 368, 985-992, 2006.\\

%\begin{thebibliography}{}
%\end{thebibliography}

\end{document}